\documentstyle[12pt]{article}

\begin{document}

\setcounter{page}{1}

\pagestyle{plain} \vspace{1cm}
\begin{center}
\Large{\bf Phantom-Like Behavior of a DGP-Inspired Scalar-Gauss-Bonnet Gravity}\\
\small \vspace{1cm} {\bf Kourosh
Nozari$^{a,}$\footnote{knozari@umz.ac.ir}},
\quad  {\bf Tahereh Azizi$^{a,}$\footnote{t.azizi@umz.ac.ir}}\quad and
\quad {\bf M. R. Setare$^{b,}$\footnote{rezakord@ipm.ir}}\\
\vspace{0.5cm} {\it $^{a}$Department of Physics, Faculty of Basic
Sciences,\\
University of Mazandaran,\\
P. O. Box 47416-95447, Babolsar, IRAN\\
\vspace{0.5cm} $^{b}$ Faculty of Science,  Department of Physics, University of Kurdistan,
Pasdaran Ave., Sanandaj, Iran}
\end{center} \vspace{1.5cm}
\begin{abstract}
We study the phantom-like behavior of a DGP-inspired braneworld
scenario where curvature correction on the brane is taken into
account. We include a possible modification of the induced gravity
on the brane by incorporating higher order curvature terms  of
Gauss-Bonnet type. We investigate the cosmological implications of
the model and we show that the normal branch of the scenario
self-accelerates in this modified scenario without introducing any
dark energy component. Also, a phantom-like behavior can be realized
in this model without introducing any phantom field that suffers
from serious difficulties such as violation of the null energy
condition. 
\end{abstract}
\vspace{2cm}
\newpage
\section{Introduction}
Observational data indicates that our universe is currently under
accelerating expansion [1]. It seems that some unknown energy
components ( dark energy) with negative pressure are responsible for
this late-time acceleration [2]. However, understanding the nature
of dark energy is one of the fundamental problems of modern
theoretical cosmology [3]. An alternative approach to accommodate
dark energy is modifying the general theory of relativity on large
scales. Among these theories, scalar-tensor theories [4], $f(R)$
gravity [5], DGP braneworld gravity [6] and string-inspired theories
[7] are studied extensively. One interesting model of modified
gravity is the Dvali-Gabadadze-Porrati (DGP) braneworld model, in
which our universe is a $(3+1)$-dimensional brane embedded in an
infinite Minkowski bulk. This model has a very appealing
phenomenology because it predicts that $4$D Newtonian gravity on the
brane is recovered at distances shorter than a given crossover
distance $r_{c}$, whereas at scale higher than $r_{c}$ the $5$D
effects become important due to leakage of gravity into the bulk. On
the other hand, this model can explain the late-time acceleration of
the universe in its self-accelerating branch without need to
introducing a dark energy term [8]. Unfortunately, this branch
suffers from some instabilities such as appearance of ghost degrees
of freedom [9].

An alternative class of modified gravity models is the family of the
string-inspired gravities by considering additional curvature
invariant terms such as the Gauss-Bonnet (GB) term [10]. GB term
arises naturally as the leading order of the $\alpha'$ expansion of
the heterotic string theory where, $\alpha'$ is the inverse string
tension [11]. Nojiri {\it et al} [12] showed that a particular dark
energy solution can be obtained from scalar-Gauss-Bonnet cosmology.
There is another version of the Gauss-Bonnet gravity namely the
modified GB or $F(G)$ theory [13] that can also play the role of
gravitational dark energy.

One of the essential tools for understanding the nature of dark
energy is to detect the evolution of its equation of state parameter
$\omega=P/\rho$, where $P$ and $\rho$ are pressure and energy
density of dark energy component respectively. Recent analysis on
the data from the Supernova, cosmic microwave background (CMB) and
large scale structure (LSS) show that the cosmological constant fits
well to the data [14], however current data also mildly favor an
evolving dark energy with an equation of state $\omega$ larger than
$-1$ in the past and less than $-1$ today [15] with evolution across
$\omega=-1$ line in the intermediate redshift. If such a result
holds on with a large number of the observational data, this would
be a great challenge to the current models of dark energy. Firstly,
the cosmological constant [16] as a candidate for dark energy will
be excluded and dark energy must be dynamical ( note also that the
cosmological constant needs a huge amount of fine-tuning). Secondly,
the case with $\omega$ less than $-1$ which is often dubbed as
phantom dark energy [17] introduces new theoretical facilities and
challenges in this field. Phantom fields are a sort of scalar fields
with negative sign for the kinetic energy term. In fact, phantom
fields suffer from instabilities due to violation of the null energy
condition, and a phantom universe eventually ends up with a Big Rip
singularity [18]. Thus it seems natural to seek alternative
approaches to realize a phantom-like behavior without introducing
any phantom field in the model. Phantom-like behavior is the growth
of the effective dark energy density with cosmic time and in the
same time, the effective equation of state parameter should stay
always less than $-1$. In this regard, it has been shown that the
normal, non-self-accelerating branch of the DGP scenario has the
potential to explain phantom-like behavior without introducing any
phantom fields on the brane ( for a number of attempts in this
direction see [19]). The phantom-like behavior of $4$-dimensional
$f(R)$ gravity is studied in [20].

With these preliminaries, the purpose of this paper is to construct
a class of DGP-inspired braneworld scenarios that curvature
corrections are taken into account via incorporation of a
Gauss-Bonnet type invariant term in the brane part of the action.
Actually, the Gauss-Bonnet invariant should be considered in the
bulk action. However, our goal here is to consider possible
modification of the induced gravity on the brane by incorporation of
higher order curvature terms such as the Gauss-Bonnet type terms. In
fact, since Gauss-Bonnet term is an invariant in $4$-dimensions, it
takes contribution in the field equations if there is a coupling
between scalar degrees of freedom on the brane and this invariant
term. This is the main reason for incorporation of the $\phi$ field
in $f(G,\phi)$. In the previous works done by other authors in this
subject, generally an effective phantom phase is realized by
adopting a phantom ansatz for scale factor. Here as we will show,
the phantom phase is realized on the normal branch of the
DGP-inspired model with curvature correction on the brane. We note
also that generally one can show, by re-construction method, that
$f(R)$ gravity is equivalent to general relativity plus the scalar
field [20]. We investigate the effect of higher order induced
curvature correction in two alternative approaches:
Scalar-Gauss-Bonnet gravity with $F(G,\phi)$ term in the brane
action and modified Gauss-Bonnet gravity with $f(G)$ term that $G$
is the Gauss-Bonnet invariant.

The paper is organized as follows: In section $2$ we start from the
action of the DGP-inspired $F(G,\phi)$ scenario and derive the
corresponding field equations. Then, we study the cosmological
implications of this model. We show that in the presence of the
curvature correction due to a Scalar-Gauss-Bonnet term in the
action, the normal branch of the scenario has a self-accelerating
behavior. The expansion history of this model will be studied and a
comparison between luminosity distances in this model, the
$\Lambda$DGP and $\Lambda$CDM will be performed. On the other hand,
in this scenario we obtain a phantom-like behavior without
introducing a phantom field neither in the bulk nor on the brane. In
section $3$ we consider a modified Gauss-Bonnet gravity in the DGP
braneworld and we obtain the cosmological equations of the scenario.
We show that this scenario can also realize a phantom-like behavior
by adopting viable ansatz for the scale factor in the appropriate
subspaces of the model parameter space. Finally, our summery and
conclusions are presented in section $4$.

\section{A DGP-Inspired $F(G,\phi)$ Scenario}
We start with the action of a DGP-inspired $F(G,\phi)$ scenario as
follows
\begin{equation}
{\cal{S}}=\int_{bulk}d^{5}X\sqrt{-{}^{(5)}g}\frac{m_{5}^{3}}{2}
{\cal{R}}+\int_{brane}d^{4}x\sqrt{-g}\bigg[\frac{m_{4}^{2}}{2}R+m_{5}^{3}
\overline{K}+F(G,\phi)+{\cal{L}}_{m}\bigg].
\end{equation}
Here $X^{A}$ with $A=0,1,2,3,5$ are coordinates in the bulk, while
$x^{\mu}$ with $\mu=0,1,2,3$ are induced coordinates on the brane.
$m_{5}^{3}$ is the 5-dimensional Planck mass and ${\cal{R}}$ is the
$5$-dimensional Ricci scalar. Also,  $m_{4}^{2}$ is the
$4$-dimensional Planck mass and $\overline{K}$ is the trace of the
mean extrinsic curvature on the brane in the higher dimensional
bulk, corresponding to the York-Gibbons-Hawking boundary term [21]
defined as
\begin{equation}
\overline{K}_{\mu\nu}=\frac{1}{2}\lim_{\epsilon\rightarrow+0}
\bigg(\Big[K_{\mu\nu}\Big]_{y=-\epsilon}+\Big[K_{\mu\nu}\Big]_{y=+\epsilon}\bigg).
\end{equation}
$y$ is the coordinate of the extra dimension and the brane is
located at $y=0$. $R$ is the induced Ricci scalar on the brane, and
the \emph{Scalar-Gauss-Bonnet} term is defined as
$$F(G,\phi)\equiv-\frac{1}{2}\partial_{\mu}\phi\partial^{\mu}\phi-V(\phi)+f(\phi)G(R).$$
By definition, the Gauss-Bonnet invariant $G(R)$ is given by
\begin{equation}
G(R)=R^2-4R_{\mu\nu}R^{\mu\nu}+R_{\mu\nu\alpha\beta}R^{\mu\nu\alpha\beta},
\end{equation}
where all terms on the right hand side are functions of the induced
curvature on the brane. In the action (1), ${\cal{L}}_{m}$ is the
Lagrangian of the other matters localized on the brane. It is
important to note that in the matter action, the matter is minimally
coupled to the metric and not to the scalar field, making the
Gauss-Bonnet gravity a metric theory which leads to conservation of
matter on the brane. Note also that we have incorporated possible
modification of the induced gravity on the brane by inclusion of the
higher order induced curvature terms via the Gauss-Bonnet invariant.
The corresponding energy-momentum tensor $T_{\mu\nu}^{(m)}$ is
defined as
\begin{equation}
T_{\mu\nu}^{(m)}=-\frac{2}{\sqrt{-g}}\frac{
\delta{(\sqrt{-g}\cal{L}}_{m})}{\delta g^{\mu\nu}}.
\end{equation}
Variation of the action (1) with respect to the scalar field, gives
the equation of motion for the scalar field on the brane,
\begin{equation}
\nabla^{2}\phi-V'(\phi)+f'(\phi)G(R)=0,
\end{equation}
where the prime denotes a derivative with respect to the scalar
field $\phi$. On the other hand, the bulk-brane Einstein's equations
calculated from the action (1) are given by
$$m^{3}_{5}\left({\cal R}_{AB}-\frac{1}{2}g_{AB}{\cal
R}\right)+
m^{2}_{4}{\delta_{A}}^{\mu}{\delta_{B}}^{\nu}\bigg[R_{\mu\nu}-
\frac{1}{2}g_{\mu\nu}R-\frac{1}{2}g^{\mu\nu}f(\phi)G(R)
+2f(\phi)RR^{\mu\nu}$$
\begin{equation}
+4f(\phi)R^{\mu}\\_{\rho}R^{\nu\rho}
+2f(\phi)R^{\mu\rho\sigma\tau}R^{\mu}\\_{\rho\sigma\tau}-4f(\phi)R^{\mu\rho\sigma\nu}
R_{\rho\sigma}\bigg]\delta(y)
={\delta_{A}}^{\mu}{\delta_{B}}^{\nu}\tau_{\mu\nu}\delta(y),
\end{equation}
where
\begin{equation}
\tau^{\mu\nu}\equiv
T^{\mu\nu}_{(M)}+T^{\mu\nu}_{(\phi)}+T^{\mu\nu}_{(c)}.
\end{equation}
The energy-momentum tensor corresponding to the scalar field and
curvature are defined as [12]
\begin{equation}
T^{\mu\nu}_{(\phi)}=\frac{1}{2}\partial^{\mu}\phi\partial^{\nu}\phi-
\frac{1}{4}g^{\mu\nu}\partial_{\rho}\phi\partial^{\rho}\phi
-\frac{1}{2}g^{\mu\nu}V(\phi),
\end{equation}
and
$$T^{\mu\nu}_{(c)}=2[\nabla^{\mu}\nabla^{\nu}f(\phi)]R-2g^{\mu\nu}[\nabla^{2}f(\phi)]R
-4[\nabla_{\rho}\nabla^{\mu}f(\phi)]R^{\nu\rho}$$
\begin{equation}
-4[\nabla_{\rho}\nabla^{\nu}f(\phi)]R^{\mu\rho}+
4[\nabla^{2}f(\phi)]R^{\mu\nu}+4g^{\mu\nu}[\nabla_{\rho}\nabla_{\sigma}f(\phi)]R^{\rho\sigma}
-4\nabla_{\rho}\nabla_{\sigma}f(\phi)]R^{\mu\rho\sigma\nu}
\end{equation}
respectively. Equation (6) can be rewritten as follows
\begin{equation}
m^{3}_{4}\left({\cal R}_{AB}-\frac{1}{2}g_{AB}{\cal R}\right)+
m^{2}_{3}{\delta_{A}}^{\mu}{\delta_{B}}^{\nu}\left(R_{\mu\nu}-
\frac{1}{2}g_{\mu\nu}R\right)\delta(y)=
{\delta_{A}}^{\mu}{\delta_{B}}^{\nu}{\cal{T}}_{\mu\nu}\delta(y)
\end{equation}
where ${\cal{T}}_{\mu\nu}$ is the total energy-momentum on the brane
defined as
\begin{equation}
{\cal{T}}_{\mu\nu}=-\frac{1}{2}g^{\mu\nu}f(\phi)G(R)+2f(\phi)RR^{\mu\nu}+4f(\phi)R^{\mu}\\_{\rho}R^{\nu\rho}
+2f(\phi)R^{\mu\rho\sigma\tau}R^{\mu}\\_{\rho\sigma\tau}-4f(\phi)R^{\mu\rho\sigma\nu}
R_{\rho\sigma}+\tau_{\mu\nu}.
\end{equation}
From equation (10) we find
\begin{equation}
G_{AB}={\cal R}_{AB}-\frac{1}{2}g_{AB}{\cal R}=0,
\end{equation}
and
\begin{equation}
G_{\mu\nu}=R_{\mu\nu}-
\frac{1}{2}g_{\mu\nu}R=\frac{1}{m^{2}_{4}}{\cal T}_{\mu\nu}.
\end{equation}
for bulk and brane respectively. The corresponding junction
conditions relating the extrinsic curvature to the energy-momentum
tensor of the brane, have the following form
\begin{equation}
\lim_{\epsilon\rightarrow+0}
\Big[K_{\mu\nu}\Big]_{y=-\epsilon}^{y=+\epsilon}=\frac{1}{m_{5}^{3}}\Big[{\cal
T}_{\mu\nu}- \frac{1}{3}g_{\mu\nu}g^{\alpha\beta}{\cal
T}_{\alpha\beta}\Big]_{y=0}-\frac{m_{4}^{2}}{m_{5}^{3}}\Big[R_{\mu\nu}-
\frac{1}{6}g_{\mu\nu}g^{\alpha\beta}R_{\alpha\beta}\Big]_{y=0}.
\end{equation}
Using the $5$D Codacci equation, the field equation of the bulk and
the junction condition at the brane lead to the conservation of the
total energy momentum tensor of the brane so that
\begin{equation}
\nabla^{\nu}{\cal T}_{\mu\nu}=0.
\end{equation}
Note that, the matter sector of the energy momentum tensor
$T_{\mu\nu}^{(m)}$ can be described by a perfect fluid with energy
density $\rho^{(m)}$ and pressure $P^{(m)}$ and satisfies the
continuity equation by virtue of the Bianchi identity
\begin{equation}
\dot{\rho}^{(m)}+3H\Big(1+\omega^{(m)}\Big)\rho^{(m)}=0.
\end{equation}

\subsection{Cosmological Implications of the model}
To study the cosmology of a homogenous and isotropic brane in this
setup, we consider the following line element
\begin{equation}
ds^{2}=g_{\mu\nu}dx^{\mu}dx^{\nu}+b^{2}(y,t)dy^{2}=-n^{2}(y,t)dt^{2}+a^{2}(y,t)\gamma
_{ij}dx^{i}dx^{j}+b^{2}(y,t)dy^{2}.
\end{equation}
In this relation, $\gamma_{ij}$ is a maximally symmetric
3-dimensional metric defined as
$\gamma_{ij}=\delta_{ij}+k\frac{x_{i}x_{j}}{1-kr^{2}}$ where
$k=-1,0,1$ corresponding to possible spatial geometries of the brane
and $r^{2}=x_{i}x^{i}$. Now, choosing a Gaussian normal coordinate
system so that $b^{2}(y,t)=1$, leads to the following form of the
junction condition (14) on the brane with the non-vanishing
components of the Einstein's tensor in the bulk
\begin{equation}
\lim_{\epsilon\rightarrow+0}
\Big[\partial_{y}n\Big]_{y=-\epsilon}^{y=+\epsilon}(t)=\frac{2nm_{4}^{2}}{m_{5}^{3}}
\Big[\frac{\ddot{a}}{n^2a}-\frac{\dot{a}^2}{2n^2a^2}-\frac{\dot{n}\dot{a}}{n^3a}-\frac{k}{2a^2}\Big]_
{y=0}+\frac{n}{3m_{5}^{3}}\Big[2\rho^{(tot)}+3P^{(tot)}\Big]_{y=0},
\end{equation}
and
\begin{equation}
\lim_{\epsilon\rightarrow+0}
\Big[\partial_{y}a\Big]_{y=-\epsilon}^{y=+\epsilon}(t)=\frac{m_{4}^{2}}{m_{5}^{3}}
\Big[\frac{\dot{a}^2}{n^2a}+\frac{k}{a}\Big]_
{y=0}-\Big[\frac{\rho^{(tot)}a}{2m_{5}^{3}}\Big]_{y=0}.
\end{equation}
With these equations, the generalized Friedmann equation for the
cosmological dynamics on the brane can be written as follows
\begin{equation}
m_{4}^{2}\bigg[H^{2}+\frac{k}{a^2}-\frac{8\pi
G\Big(\rho^{(m)}+\rho^{(GB)}\Big)}{3}\bigg]^{2}=m_{5}^{6}\bigg(H^{2}+\frac{k}{a^2}-\frac{C}{a^4}\bigg).
\end{equation}
where the last term in the right hand side of this equation
represents the dark radiation term and $C$ is an integration
constant. It is important to note that this term can be neglected at
present time due to its fast decaying behavior. The energy density
corresponding to the Gauss-Bonnet term is defined as
\begin{equation}
\rho^{(GB)}\equiv\frac{1}{2}\dot{\phi}^{2}+V(\phi)-24\dot{\phi}f'(\phi)H^3,
\end{equation}
and the corresponding pressure is defined as
\begin{equation}
P^{(GB)}=\frac{1}{2}\dot{\phi}^{2}-V(\phi)+8\frac{\partial}{\partial
t}\Big(H^2\dot{f}\Big)+16H^3\dot{\phi}f'(\phi),
\end{equation}
where a dot refers to derivative with respect to the cosmic time of
the brane. The modified Friedmann equation (20) can be rewritten as
follows
\begin{equation}
H^{2}+\frac{k}{a^2}=\frac{\rho^{(m)}+\rho^{(GB)}}{3m_{4}^{2}}
+\frac{1}{2r_{c}^{2}}+\varepsilon\sqrt{\frac{1}{4r_{c}^{4}}+\frac{1}{r_{c}^{2}}
\Big(\frac{\rho^{(m)}+\rho^{(GB)}}{3m_{4}^{2}}\Big)}.
\end{equation}
where $\varepsilon=\pm1$ is associated to two possible branches of
the scenario corresponding to two different embeddings of the brane
in the bulk and $r_{c}=\frac{m_{4}^{2}}{2m_{5}^{3}}$ is the
crossover scale. In the distance scale lower than this scale,
gravity behaves as usual $4$D scenario but in the distance scales
higher than the crossover scale, gravity leaks to the extra
dimension and this leakage leads to weakness of gravity in the large
scales, so the universe accelerates. In other words, weakness of
gravity at large scales plays the role of a negative pressure fluids
that accelerates the universe expansion. Equation (23) as the
modified Friedmann equation in this DGP-inspired $F(G,\phi)$
scenario is similar to the standard induced gravity result, but we
note that here there is an important difference which is hidden in
the definition of the total energy density and pressure on the
brane. There is a contribution from higher order curvature
corrections in these total quantities and hence in the modified
Friedmann equation (23). In another words, the effect of the higher
order curvature corrections in the brane action are interpreted as a
contribution in the total energy density and pressure on the brane.
On the other hand, the modified Raychaudhuri equation in this setup
has the following form
$$
\bigg\{1+\frac{1}{3m_{4}^{2}}r_{c}\Big[\rho^{(m)}+\rho^{(GB)}-6r_{c}\big(
H^{2}+\frac{k}{a^2}\big)\Big]\bigg\}\Big(\dot{H}-\frac{k}{a^2}\Big)
=-\frac{1}{12m_{4}^{2}}\Big(\rho^{(m)}+\rho^{(GB)}+P^{(m)}+P^{(GB)}\Big)
$$
\begin{equation}
\times\bigg[\rho^{(m)}+\rho^{(GB)}-6r_{c}\Big(
H^{2}+\frac{k}{a^2}\Big)\bigg].
\end{equation}

\subsection{Self-accelerating the normal branch with curvature correction}

In which follows we assume a spatially flat FRW brane. In this case,
we can rewrite the modified Friedmann equation (23) in the following
form
\begin{equation}
H^{2}=\frac{8\pi G}{3}\Big(\rho^{(m)}+\rho^{(GB)}\Big)+
\varepsilon\frac{H}{r_{c}}.
\end{equation}
where $G\equiv1/8\pi m_{4}^{2}$ is the $4$D gravitational constant.
One of the simplest cosmological model that can exhibit acceleration
of the universe is a de Sitter spacetime. So, it seems interesting
to looking for such solutions in our model. With this motivation, we
can probe the self-accelerating solution of the modified normal
branch of this scenario. Assuming a de Sitter spacetime, i.e.,
looking for solutions of the form $a(t)= a_{0} \exp (H_{0}t)$, where
$H_{0}$ and $a_{0}$ are constants, the modified Friedmann equation
(25) in a de Sitter universe with Hubble rate $H_{0}$ is given by
\begin{equation}
\rho^{(m)}+V_{0}=constant.
\end{equation}
where the subscript $0$ stands for quantities evaluated in the de
Sitter phase and the scalar field is assumed to be constant in time.
This relation shows that the matter energy density is constant by
the virtue of the continuity equation (16), leads to a cosmological
constant behavior for the matter sector ($\omega^{(m)}=-1$) on the
brane. In which follows, we neglect the matter content of the brane
in our analysis of de Sitter brane. Using equation (23), the Hubble
rate can be expressed as follows
\begin{equation}
2r_{c}^{2}H_{0}^{2}=1+\frac{16\pi
Gr_{c}^{2}V_{0}}{3}+\varepsilon\sqrt{1+\frac{32\pi
Gr_{c}^{2}}{3}V_{0}}.
\end{equation}
Note that the DGP braches can be recovered by imposing $V_{0}=0$.
Indeed, the case $\varepsilon=1$ branch is corresponding to the de
Sitter self-accelerating solution, and the $\varepsilon=-1$ case
corresponds to the normal branch leads to a Minkowski spacetime or
the non-self-accelerating solution. In the presence of the curvature
correction due to a Scalar-Gauss-Bonnet term in the action, the
normal branch of the scenario self-accelerates too. It is important
to note that the normal branch has not the difficulties of the
self-accelerating one such as the ghost instabilities. In figure
$1$, we plot the behavior of the re-scaled squared Hubble parameter
$2r_{c}^{2}H_{0}^{2}$ versus the re-scaled energy density
$\frac{16\pi Gr_{c}^{2}}{3}V_{0}$ for two branches of the solutions.
This figure shows that the normal branch of the scenario exhibits a
self-accelerating behavior because of the presence of an extra
energy density originated from curvature corrections.
\begin{figure}[htp]
\begin{center}
\includegraphics{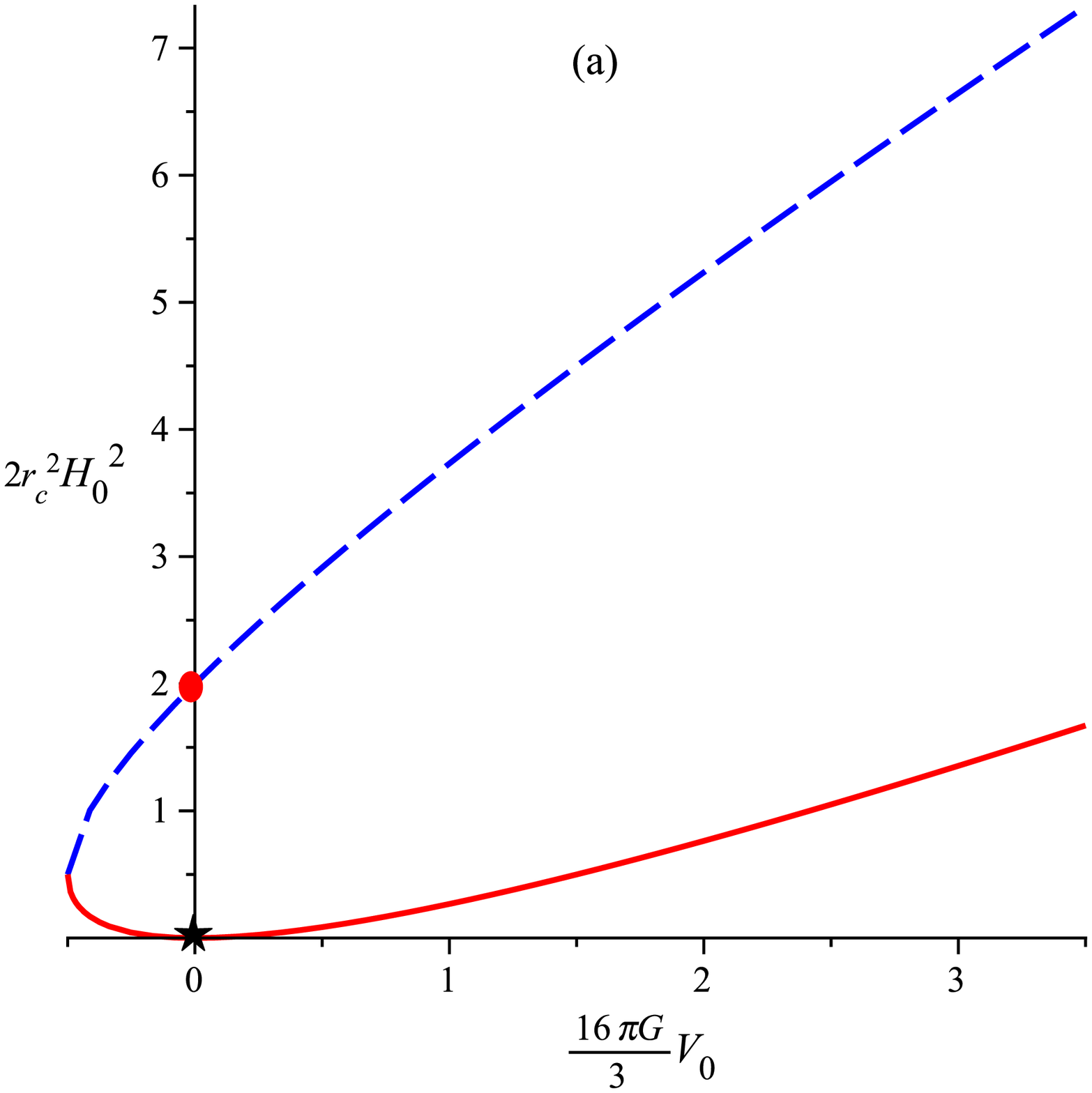} \vspace{1cm}\includegraphics{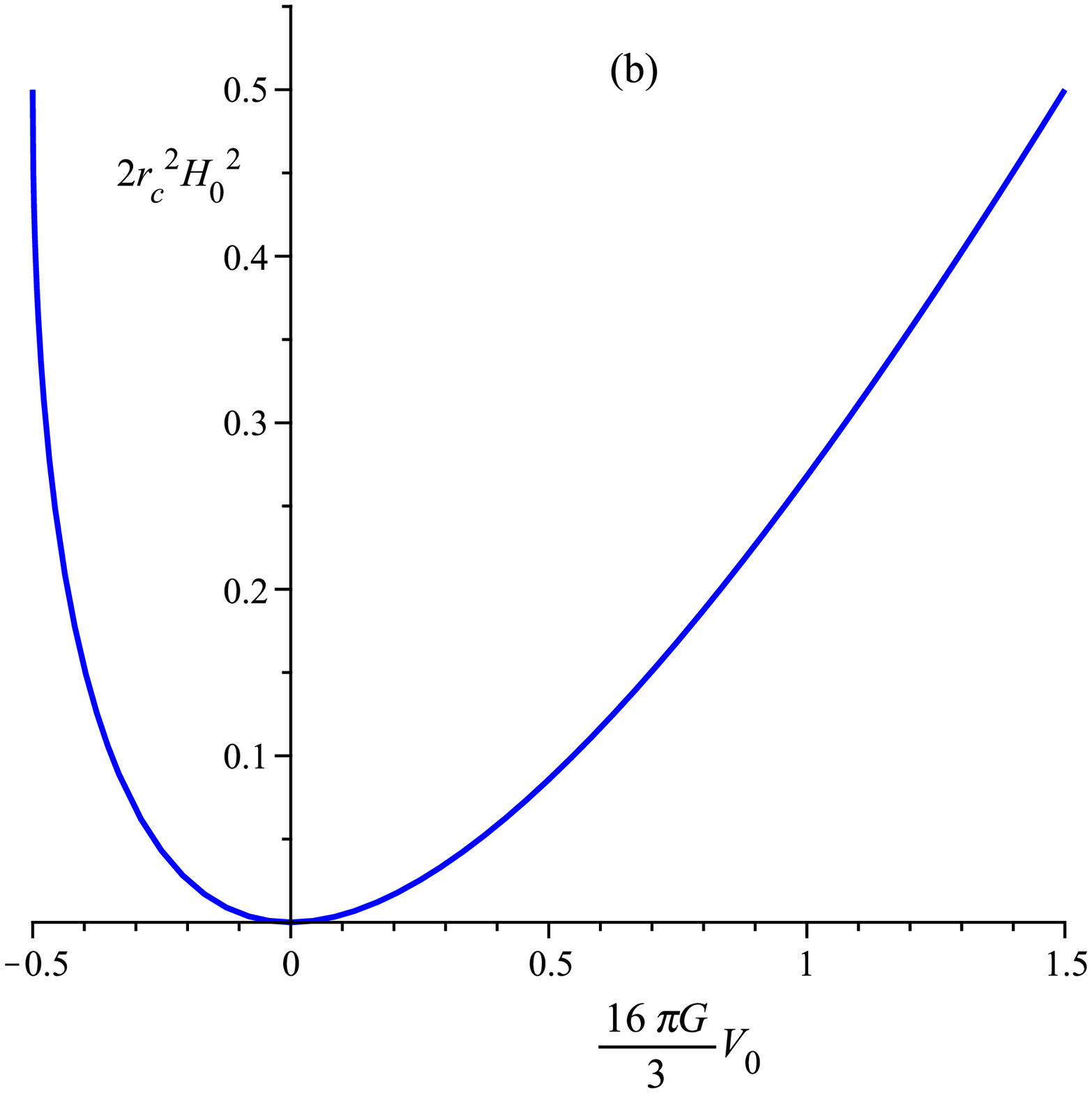}\vspace{3cm}
\end{center}
\vspace{2.5cm}
 \caption{\small {(a)\, Behavior of the
re-scaled squared Hubble parameter $2r_{c}^{2}H_{0}^{2}$ versus the
re-scaled energy density $\frac{16\pi Gr_{c}^{2}}{3}V_{0}$ for two
branches of the scenario. The solid curve (with star) is
corresponding to the normal branch and the dashed curve ( with
circle) corresponds to the self-accelerating branch of the scenario.
(b) A zoom on the normal branch of the DGP scenario with curvature
effect. }}
\end{figure}

Up to this point, we have focused on the effects of a curvature
correction term on the DGP model. Now we consider the effect of
higher order embedding on the $4$D Scalar-Gauss-Bonnet model. In
other words, we study the effect of extra dimension on the
cosmological dynamics on the brane. The Hubble parameter (27) can be
expressed in a $4$D regime as follows
\begin{equation}
H_{0}^{2}=H_{(4)}^{2}+\frac{1+\varepsilon\sqrt{1+\frac{32\pi
Gr_{c}^{2}}{3}V_{0}}}{2r_{c}^{2}}.
\end{equation}
where $H_{(4)}^{2}=\frac{8\pi G}{3}V_{0}$. This relation shows that
the effect of an extra dimension is a shift on the $4$D Hubble rate.
In order to de Sitter brane to be close to the standard $4$D regime
( with $H_{(0)}^{2}\sim H_{(4)}^{2}$), the following condition
should be satisfied
\begin{equation}
\Bigg|\frac{3\bigg(1+\varepsilon\sqrt{1+\frac{32\pi
Gr_{c}^{2}}{3}V_{0}}\bigg)}{16\pi Gr_{c}^{2}V_{0}}\Bigg|\ll1.
\end{equation}

\subsection{Phantom-Like Mimicry}
We can express the Friedmann equation (23) for the normal branch of
this DGP-inspired scenario in a dimensionless form as follows

 $$E^{2}=\frac{ H^{2}(z)}{
 H_{0}^{2}}=\Omega_{m}(1+z)^{3}+\Omega_{GB}(1+z)^{3(1+\omega_{GB})}+2\Omega_{r_{c}}$$
\begin{equation}
 - 2\sqrt{\Omega_{r_{c}}}\sqrt{\Omega_{m}(1+z)^{3}+\Omega_{GB}(1+z)^{3(1+\omega_{GB})}+
 \Omega_{r_{c}}}
\end{equation}
where $\Omega_{m}=\frac{8\pi G\rho_{0}^{(m)}}{3H_{0}^{2}}$\, ,\,
$\Omega_{GB}=\frac{8\pi G\rho_{0}^{(GB)}}{3H_{0}^{2}}$
   and
  $\Omega_{r_{c}}=\frac{1}{4r_{c}^{2}H_{0}^{2}}$\,. Now we study the
constraints imposed on this model. At redshift $z=0$, equation (30)
can be expressed as ( see Ref. [22] for a general framework in this
direction)
\begin{equation}
1=\bigg(\sqrt{\Omega_{m}+\Omega_{GB}+\Omega_{r_{c}}}-\sqrt{\Omega_{r_{c}}}\bigg)^{2}.
\end{equation}
This expression indicates that $\Omega_{r_{c}}$ should be positive
and $\Omega_{m}+\Omega_{GB}+\Omega_{r_{c}}>0$ \, too. Taking the
square root of the last equation, there are two possibility for the
sign of the quantity $\Omega_{m}+\Omega_{GB}$ as follows
\begin{equation}
\Omega_{m}+\Omega_{GB}\geq 0,
\end{equation}
and
\begin{equation}
-\Omega_{r_{c}}\leq\Omega_{m}+\Omega_{GB}<0.
\end{equation}
In the first case, taking the square root of the equation (31)
yields
\begin{equation}
1+\sqrt{\Omega_{r_{c}}}=
\sqrt{\Omega_{m}+\Omega_{GB}+\Omega_{r_{c}}}\,\,.
\end{equation}
Now by squaring both sides of this equation we find
\begin{equation}
\Omega_{m}+\Omega_{GB}-2\sqrt{\Omega_{r_{c}}}=1.
\end{equation}
Similarly, for the second case the constraint equation is
\begin{equation}
\Omega_{m}+\Omega_{GB}+2\sqrt{\Omega_{r_{c}}}=1.
\end{equation}
Note that the general relativistic limit can be recovered if we set
$\Omega_{r_{c}}=0$ (or $m_{5}=0$). In this case equation (31)
implies that $\Omega_{m}+\Omega_{GB}=1$.

For a comparison between this DGP-inspired $F(G,\phi)$ scenario and
the well-known $\Lambda$DGP ( see for instance [23]) and
$\Lambda$CDM models, we study the expansion histories of these
alternative scenarios. We focus on the variation of the Luminosity
Distance versus the redshift in these scenarios. The luminosity
distance for a spatially flat universe is expressed as follows
\begin{equation}
 d_{L}(z)=(1+z)\int_{0}^{z}\frac{dz}{H(z)},
\end{equation}
where for the DGP-inspired $F(G,\phi)$ scenario, the Hubble rate is
given by equation (30). For $\Lambda$DGP model, $H(z)$ is defined as
[23]
\begin{equation}
\frac{H(z)}{H_{0}}=\frac{1}{2}\Bigg[-\frac{1}{r_{0}H_{0}}+
\sqrt{\Big(2+\frac{1}{r_{0}H_{0}}\Big)^{2}+4\Omega_{M}^{0}
\Big[(1+z)^{3}-1\Big]}\,\,\Bigg].
\end{equation}
The luminosity distance of the $\Lambda$CDM model is given by
\begin{equation}
 d_{L}^{\Lambda CDM}(z)=(1+z)\int_{0}^{z}\frac{H_{0}dz}{\sqrt{\Omega_{m}(1+z)^{3}+\Omega_{\Lambda}}}.
\end{equation}
In figure (2) we show that the luminosity distance of the
DGP-inspired scenario with curvature correction is closer to the
$\Lambda$CDM model than the $\Lambda$DGP scenario. We note that this
result is dependent on the choice of the model parameter space.
Nevertheless, since the $\Lambda$CDM has a very good agreement with
observational data, our analysis shows that with a suitable choice
of the parameter space of the model, the DGP-inspired $F(G,\phi)$
scenario has better agreement with recent observation than
$\Lambda$DGP. This result seems to be reasonable since the
DGP-inspired $F(G,\phi)$ scenario has wider parameter space than the
$\Lambda$DGP model.
\begin{figure}[htp]
\begin{center}\includegraphics{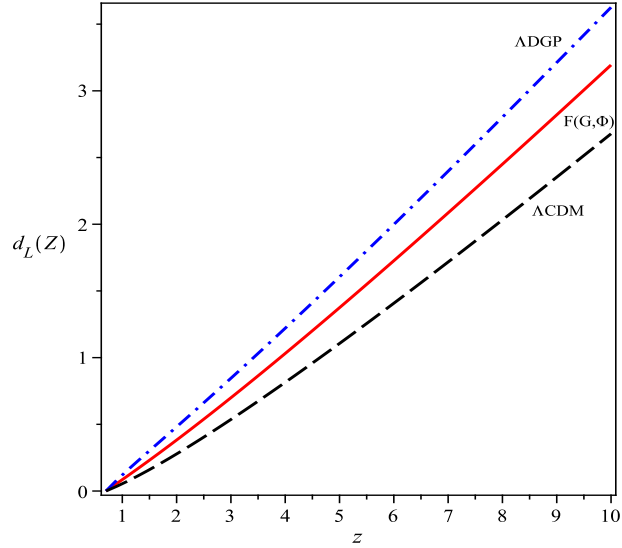} \vspace{7cm}
\end{center}
 \caption{\small {Luminosity distance versus the redshift for three
 alternative scenarios. In this figure we have set $\Omega_{m}=0.25$, $\Omega_{rc}=0.12$,
 and $\Omega_{GB}=-0.56$.
 With these values, we calculate $\Omega_{GB}$ using the constraint
 equation (35). There is better agreement between DGP-inspired $F(G,\phi)$ scenario and the $\Lambda$CDM
 at least in the parameter values adopted here. }}
\end{figure}

Now, we show that our model can lead to the phantom-like behavior on
the brane without need to introduce any kind of the phantom scalar
fields that violate the null energy condition. For this goal, the
standard Friedman equation can be written as an effective form
\begin{equation}
H^{2}=\frac{8\pi G}{3}\Big(\rho^{(m)}+\rho^{(DE)}_{eff}\Big)
\end{equation}
where $\rho^{(m)}$ is the energy density of the standard matter and
$\rho^{(DE)}_{eff}$  is the energy density corresponding to dark
energy. Comparison between the normal branch of equation (20) with
this equation leads to the following relation for
$\rho^{(DE)}_{eff}$
\begin{equation}
\frac{8\pi
G}{3}\rho^{(DE)}_{eff}=\rho^{(GB)}+\frac{3}{2r_{c}^{2}}\Bigg[1-\sqrt{1+\frac{32\pi
Gr_{c}^{2}}{3} \Big(\rho^{(m)}+\rho^{(GB)}\Big)}\Bigg]
\end{equation}
where we have set $8\pi G=1$. With the observable quantities, this
equation is given by
\begin{equation}
  \frac{\rho^{(DE)}_{eff}}{H_{0}^{2}}=\Omega_{GB}(1+z)^{3(1+\omega_{GB})}+6\Omega_{r_{c}}-
 6\sqrt{\Omega_{r_{c}}}\sqrt{\Omega_{m}(1+z)^{3}+\Omega_{GB}(1+z)^{3(1+\omega_{GB})}+
 \Omega_{r_{c}}}
\end{equation}
Note that $\rho^{(DE)}_{eff}$ and
$\omega_{eff}=\frac{P_{eff}}{\rho_{eff}}$ satisfy the continuity
equation in the same way as the general relativity
\begin{equation}
\dot{\rho}_{eff}^{(DE)}+3H(1+\omega_{eff})\rho_{eff}^{(DE)}=0.
\end{equation}
In this respect, the effective equation of state of dark energy can
be expressed as follows
\begin{equation}
\omega_{eff}=-1+\frac{(1-\sqrt{\Omega_{r_{c}}})\Omega_{GB}(1+z)^{3(1+\omega_{GB})}-
\sqrt{\Omega_{r_{c}}}\Omega_{m}(1+z)^{3}}
{\rho^{(DE)}_{eff}\sqrt{\Omega_{m}(1+z)^{3}+\Omega_{GB}(1+z)^{3(1+\omega_{GB})}+
 \Omega_{r_{c}}}}
\end{equation}
In figure $3$ we plot the effective energy density and equation of
state parameter of the model versus the redshift $z$. This figure
shows a growing behavior of the effective energy density with
decreasing of $z$; a requirement for phantom-like prescription. On
the other hand, the effective equation of state parameter remains
less than $-1$ in the phantom phase at future. Especially, the
figure shows that the effective equation of state parameter has
passed the phantom divide line $\omega_{eff}=-1$ in the near past.
\begin{figure}[htp]
\begin{center}
\includegraphics{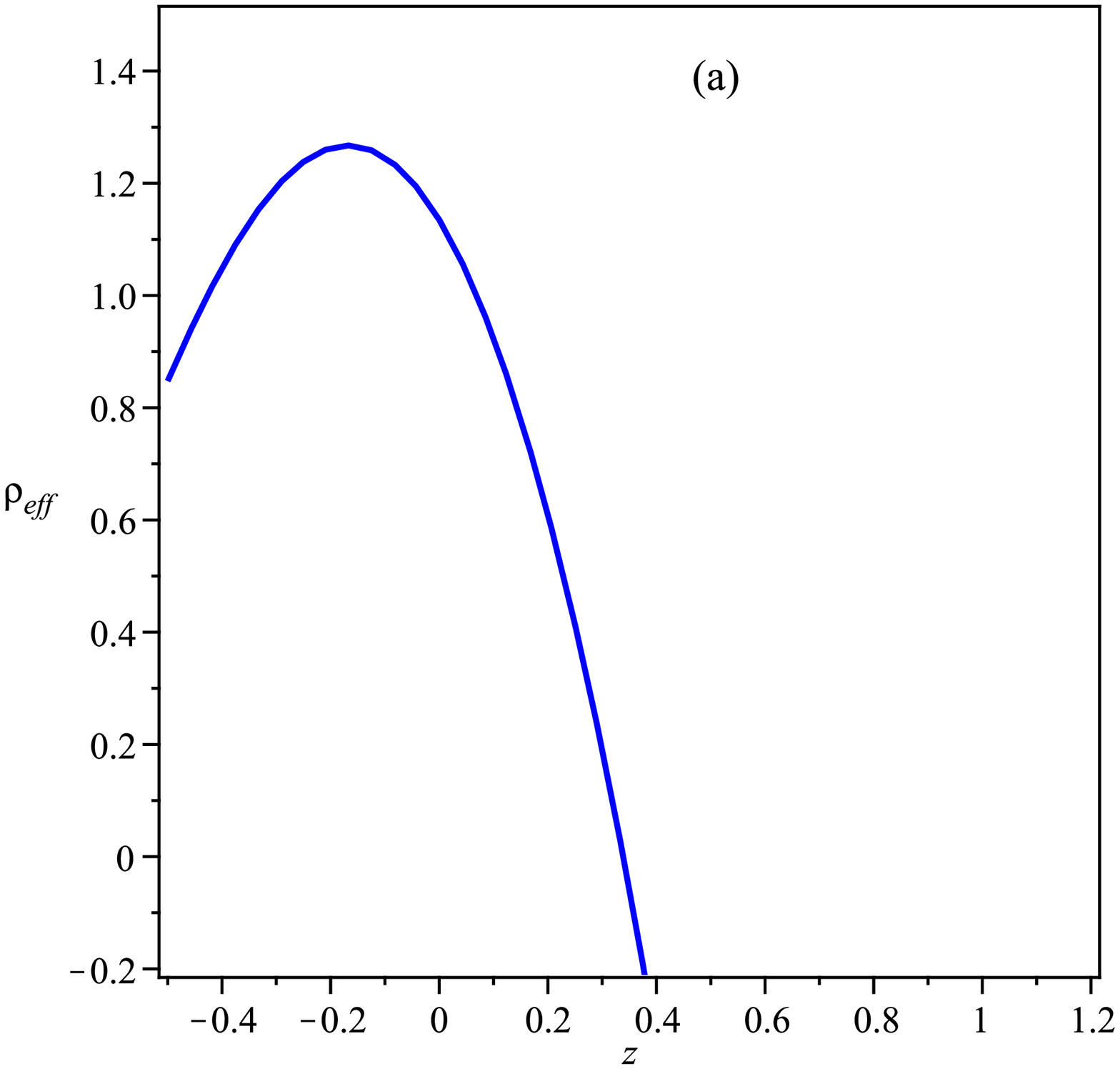} \vspace{1cm}\includegraphics{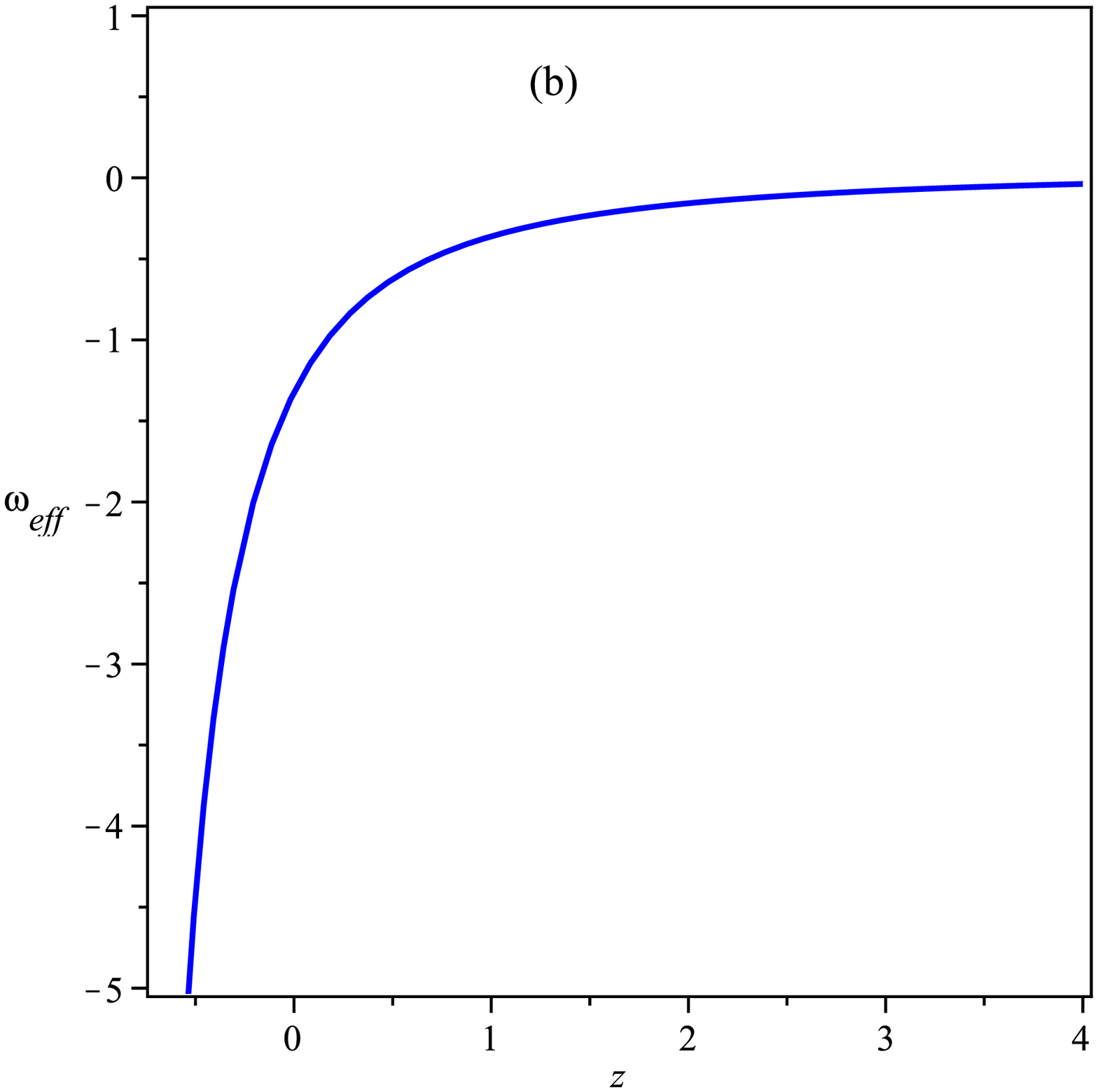}\vspace{5cm}
\end{center}
\vspace{1cm} \caption{\small { a) Variation of the effective dark
energy density versus the redshift. The effective dark energy
density grows by decreasing $z$ and therefore shows a phantom-like
behavior. b) The effective equation of state parameter versus the
redshift. The braneworld universe has entered in the phantom phase
in the near past and there is a smooth crossing of the phantom
divide line in this setup. }}
\end{figure}
From figure $4$ we can obtain appropriate ranges of $z$ to fulfill
the null energy condition,
($\rho_{eff}^{(DE)}+P_{eff}^{(DE)}\geq0$). Clearly, this condition
is satisfied in the present and future times at least in some
subspaces of the model parameter space. As a result, the
phantom-like prescription is realized in this DGP-inspired scenario
without introduction of any phantom field neither in the bulk nor on
the brane.
\begin{figure}[htp]
\begin{center}\includegraphics{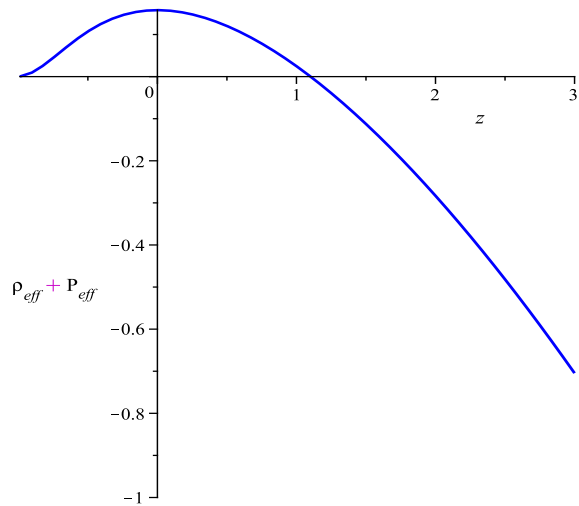} \vspace{6cm}
\end{center}
 \caption{\small {The null energy condition is fulfilled for $z<1.1$\,\,. }}
\end{figure}
\newpage
\section{An Extension with $f(G)$ Term on the Brane }
In this section we consider a modified Gauss-Bonnet (modified-GB)
term on the action of the DGP brane. The modified-GB gravity has
been extensively studied in the literature ( see for example [13] ).
Especially in Ref. [24] it has been shown that this model may play
the role of a gravitational alternative to dark energy proposal. On
the other hand, the modified-GB scenario has no ghosts and is
stable, and it also passes the solar system tests. The action of the
DGP-inspired modified Gauss-Bonnet gravity is given by
\begin{equation}
{\cal{S}}=\int_{bulk}d^{5}X\sqrt{-{}^{(5)}g}\frac{m_{5}^{3}}{2}
{\cal{R}}+\int_{brane}d^{4}x\sqrt{-g}\bigg[\frac{m_{4}^{2}}{2}R+m_{5}^{3}
\overline{K}+f(G)+{\cal{L}}_{m}\bigg].
\end{equation}
Note that the brane part of the action is dynamically equivalent to
the Scalar-Gauss-Bonnet gravity with vanishing kinetic-energy term
(see [24] for more details). In this action, $f(G)$ is an arbitrary
function of the Gauss-Bonnet invariant which is defined in equation
(3). In the same line as the previous section, the bulk-brane
Einstein's equations calculated from action (45) are given by
\begin{equation}
m^{3}_{4}\left({\cal R}_{AB}-\frac{1}{2}g_{AB}{\cal R}\right)+
m^{2}_{3}{\delta_{A}}^{\mu}{\delta_{B}}^{\nu}\left(R_{\mu\nu}-
\frac{1}{2}g_{\mu\nu}R\right)\delta(y)=
{\delta_{A}}^{\mu}{\delta_{B}}^{\nu}{\cal{T}}_{\mu\nu}\delta(y),
\end{equation}
where ${\cal{T}}_{\mu\nu}$ is the total energy-momentum on the brane
defined as ${\cal{T}}_{\mu\nu}=T_{\mu\nu}^{(m)}+T_{\mu\nu}^{(f)}$.
The energy-momentum tensor $T_{\mu\nu}^{(f)}$ is corresponding to
the Gauss-Bonnet curvature on the brane and is given by
$$
{T^{\mu\nu}}^{(f)}=\frac{1}{2}g^{\mu\nu}f(G)-2f'(G)RR^{\mu\nu}+4f'(G)R^{\mu}\\_{\rho}R^{\nu\rho}
-2f'(G)R^{\mu\rho\sigma\tau}R^{\nu}\\_{\rho\sigma\tau}-4f'(G)R^{\mu\rho\sigma\nu}
R_{\rho\sigma}$$$$+2[\nabla^{\mu}\nabla^{\nu}f'(G)]R
-2g^{\mu\nu}[\nabla^{2}f'(G)]R
-4[\nabla_{\rho}\nabla^{\mu}f'(G)]R^{\nu\rho}
-4[\nabla_{\rho}\nabla^{\nu}f'(G)]R^{\mu\rho}+
4[\nabla^{2}f'(G)]R^{\mu\nu}$$
\begin{equation}
+4g^{\mu\nu}[\nabla_{\rho}\nabla_{\sigma}f'(G)]R^{\rho\sigma}
-4\nabla_{\rho}\nabla_{\sigma}f'(G)]R^{\mu\rho\sigma\nu}\, ,
\end{equation}
where the prime denotes a derivative with respect to the $G$. From
the equation (46), the field equations can be deduced as follows
\begin{equation}
G_{AB}={\cal R}_{AB}-\frac{1}{2}g_{AB}{\cal R}=0,
\end{equation}
and
\begin{equation}
G_{\mu\nu}=R_{\mu\nu}-
\frac{1}{2}q_{\mu\nu}R=\frac{1}{m^{2}_{4}}{\cal T}_{\mu\nu}.
\end{equation}
for the bulk and brane respectively. By imposing the junction
conditions and using the FRW metric on the brane ( equations
(17)-(19) ), we find the following Friedmann equation for the normal
branch of the scenario
\begin{equation}
H^{2}=\frac{1}{3}\Big(\rho^{(m)}+\rho^{(f)}\Big)-\frac{H}{r_{c}}.
\end{equation}
Where we have set $m_{4}^{2}\equiv1$. The energy density and
pressure corresponding to the Gauss-Bonnet term can be defined as
[22]
\begin{equation}
\rho^{(f)}=Gf'(G)-f(G)-24^2H^4\Big(2\dot{H}^2+H\ddot{H}+4H^2\dot{H}\Big)f''(G)
\end{equation}
and
\begin{equation}
P^{(f)}=f(G)-24^2H^2\Big(3H^4+20H^2\dot{H}^2+6\dot{H}^3+
4H^3\ddot{H}+H^2\ddot{H}\Big)f''(G)
\end{equation}
respectively. Comparing equation (50) with the standard Friedmann
equation, we can define an effective energy density as follows
\begin{equation}
H^{2}=\frac{1}{3}\Big(\rho^{(m)}+\rho^{(DE)}_{eff}\Big)\,,
\end{equation}
where $\rho^{(m)}$ is the energy density of the standard matter and
$\rho^{(DE)}_{eff}$  corresponds to the energy density of dark
energy which is given by
\begin{equation}
\rho^{(DE)}_{eff}=\rho^{(f)}-\frac{3H}{r_{c}}.
\end{equation}
Now we assume $f(G)$ to be defined as $f(G)\equiv f_{0}|G|^{\beta}$
with constants $f_{0}$ and $\beta$. For $\beta<\frac{1}{2}$, the
$f(G)$ term becomes dominant in the small curvature regime. We note
that $\beta$ can take essentially both positive and negative signs.
Now the effective energy density can be rewritten in the following
form
\begin{equation}
\rho^{(DE)}_{eff}=f_{0}G^{\beta}\Bigg(f_{0}\beta-1-\Big(\frac{24\beta
H^2}{G}\Big)^2\Big[2\dot{H}^2+H\ddot{H}+4H^2\dot{H}\Big]\Bigg)
-\frac{3H}{r_{c}}.
\end{equation}
As usual, the effective quantities $\rho^{(DE)}_{eff}$ and
$\omega_{eff}=\frac{P_{eff}}{\rho_{eff}}$ satisfy the continuity
equation in the same way as the general relativity
\begin{equation}
\dot{\rho}_{eff}^{(DE)}+3H(1+\omega_{eff})\rho_{eff}^{(DE)}=0.
\end{equation}
Considering a power-law expansion on the brane, we set the scale
factor to be $a(t)=a_{0}t^{h_{0}}$  with $h_{0}=1.2$. This choice is
reliable from physical grounds since it is corresponding to an
accelerated expansion on the brane. Figure $5$ shows the behavior of
the effective dark energy density versus the redshift for two signs
of $\beta$. As this figure shows, although the effective energy
density increases by decreasing $z$, its value remains positive only
for negative values of $\beta$. This property is illustrated
explicitly in figure $5(b)$. In this figure we have plotted
$\rho_{eff}^{(DE)}$ versus $\beta$ for various values of the
redshift.
\begin{figure}[htp]
\begin{center}
\includegraphics{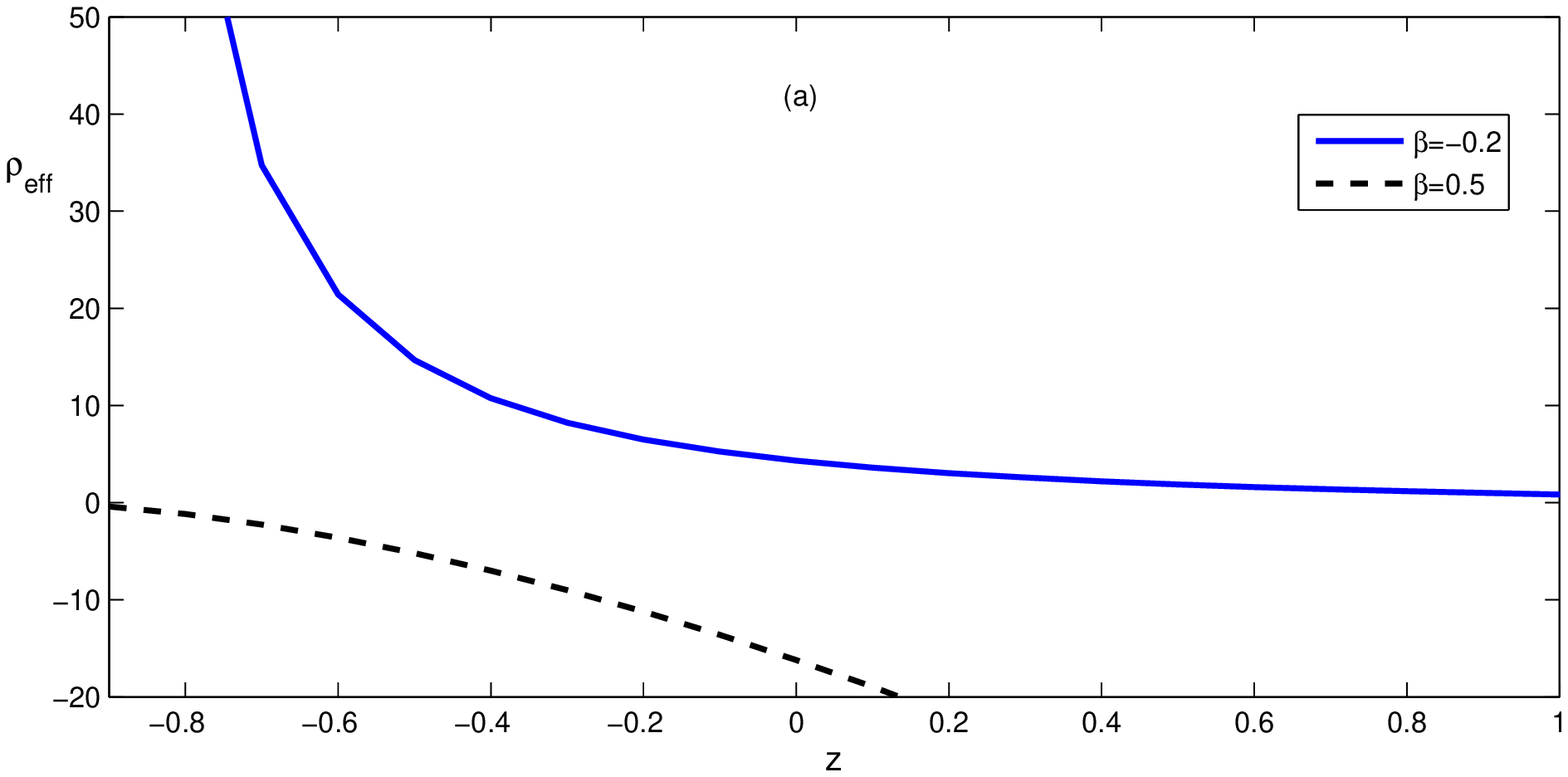} \vspace{1cm}\includegraphics{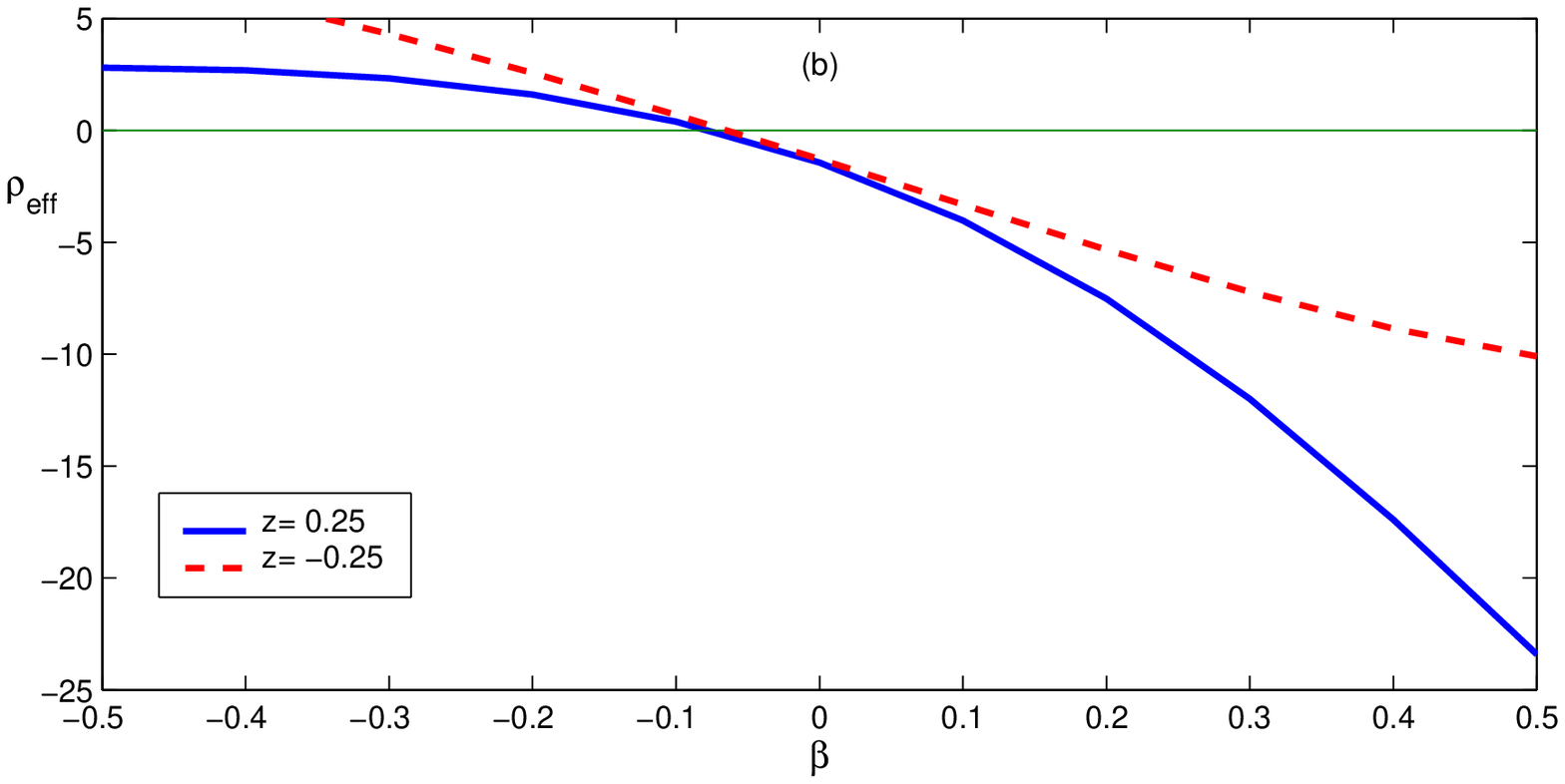}\vspace{1.5cm}
\end{center}
\vspace{2.5cm}
 \caption{\small {a) Variation of the effective energy density versus the redshift
 for $\beta=-0.2$ and $\beta=0.5$. In both cases $\rho_{eff}^{(DE)}$ increases by decreasing $z$,
   but its value always remains negative for positive values of
   $\beta$.\,\, b) Variation of $\rho_{eff}^{(DE)}$ versus $\beta$ for $z=\pm 0.25$.
   For positive values of $\beta$ the effective dark energy is always
   negative and the phantom-like prescription breaks down in this case.}}
\end{figure}
The acceptable range of $\beta$ can be deduced in a fascinating
manner via imposing the null energy condition (
$\rho_{eff}^{(DE)}+P_{eff}^{(DE)}\geq 0$) for this DGP-inspired
modified-GB scenario as we have shown in figure $6$. The null energy
condition is fulfilled for a narrow strip defined as
$-0.25\leq\beta\leq0.05$.
\begin{figure}[htp]
\begin{center}\includegraphics{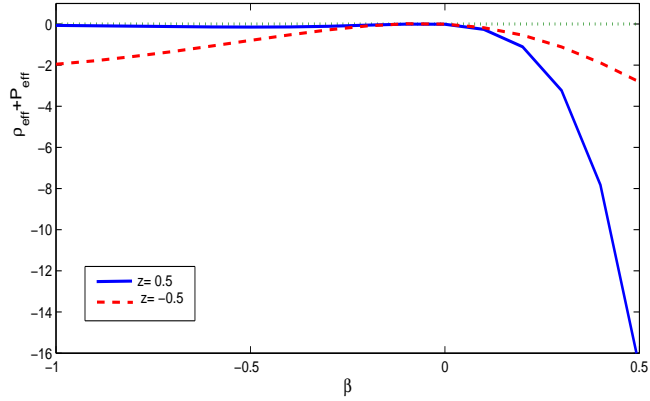} \vspace{3cm}
\end{center}
 \caption{\small {The null energy condition is fulfilled
 for a narrow strip $-0.25\leq\beta\leq0.05$. }}
\end{figure}

In figure $7(a)$, we have plotted the variation of the effective
equation of state parameter versus the redshift for $\beta=-0.2$.
Figure $7(b)$ shows that in the case of positive $\beta$,
$\omega_{eff}$ is in the quintessence phase ($\omega_{eff}>-1$). For
negative values of $\beta$, the effective equation of state lies in
the phantom phase with ($\omega_{eff}<-1$). A unified treatment of
the above results extracted from figures shows that an effective
phantom-like behavior can be realized in this setup in the region
$-0.25\leq \beta \leq0.05$. Therefore, a DGP-inspired modified
Gauss-Bonnet gravity has the potential to realize a phantom-like
behavior in appropriate subspaces of the model parameter space. It
is important to note that with scale factor defined as
$a=a_{0}(t_{s}-t)^{h_{0}}$, an effective phantom phase can be
obtained also with negative $h_{0}$ ( see for instance [24] and
references therein). However, in our setup an effective phantom
phase is realized with a positive value of $h_{0}$ by virtue of the
induced gravity effects.

\begin{figure}[htp]
\begin{center}
\includegraphics{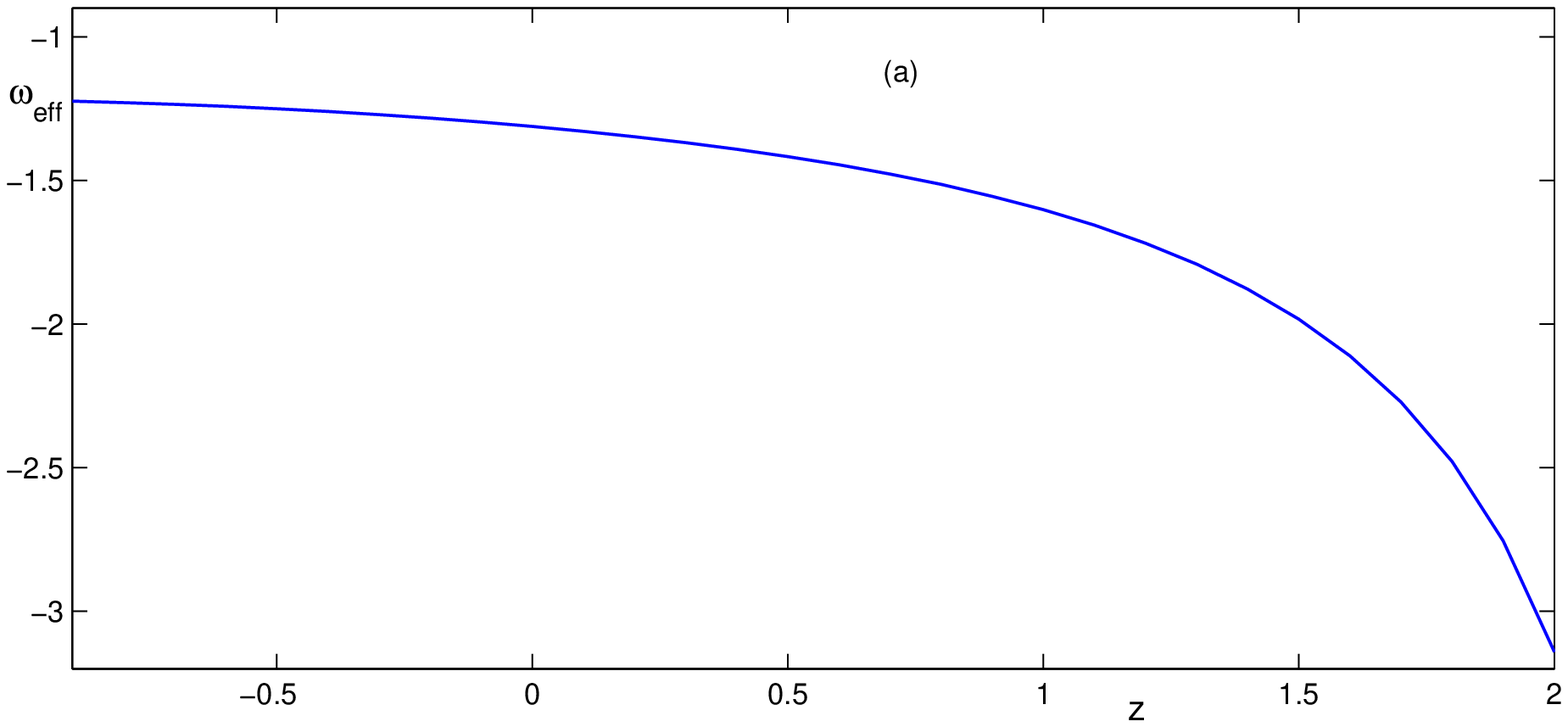} \vspace{1cm}\includegraphics{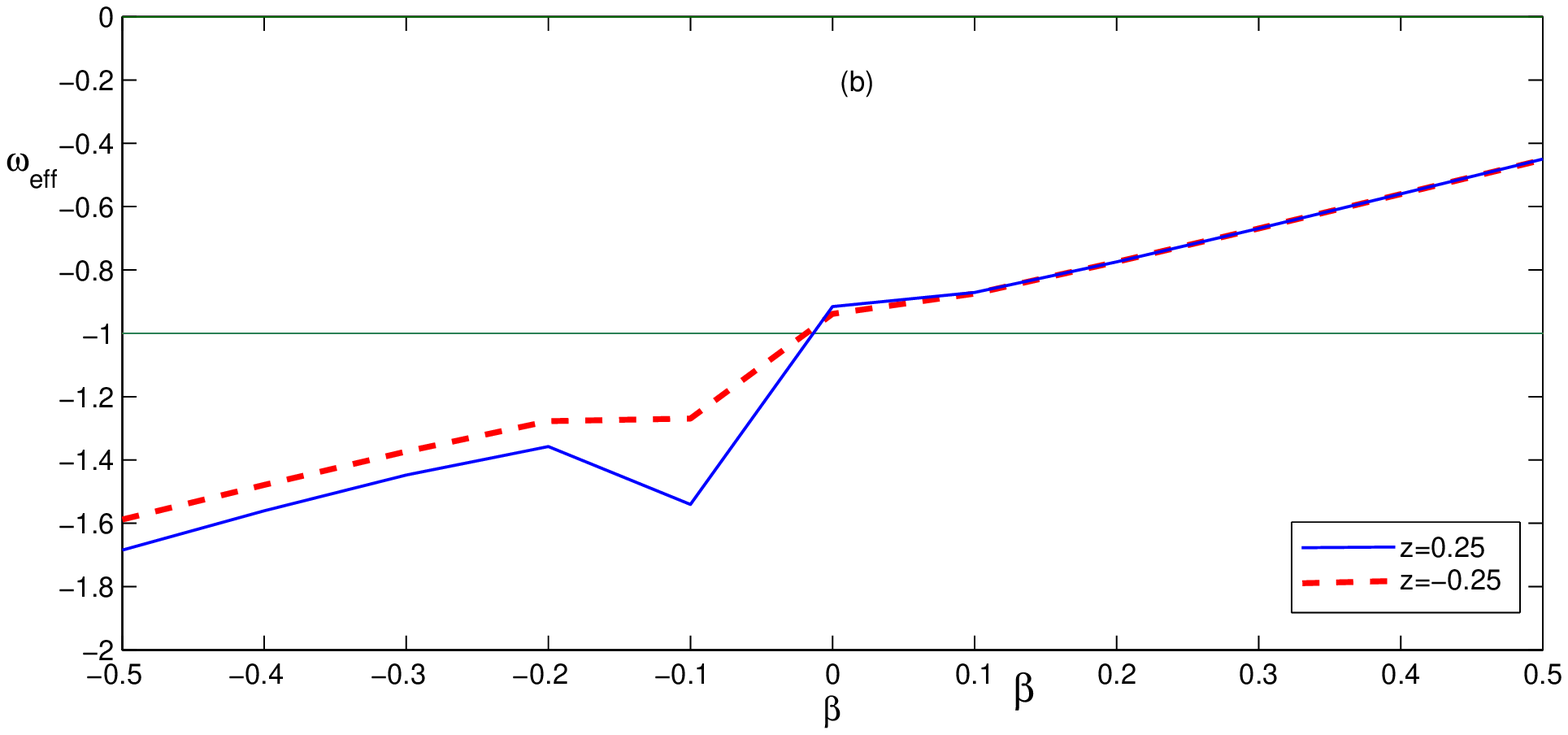}\vspace{3cm}
\end{center}
\vspace{1.5cm}
 \caption{\small {a) Variation of the effective equation of state parameter
  versus the redshift for $\beta=-0.2$ which remains in the phantom phase.
  b) Variation of the $\omega_{eff}$ versus $\beta$ for $z=\pm 0.25$. In the case of
positive $\beta$, $\omega_{eff}$ is larger than $-1$ but for
negative values of $\beta$ the effective equation of state lies in
the phantom phase with $\omega_{eff}<-1$. }}
\end{figure}
\section{Summary and Conclusion}
 In this paper we have studied possible realization of the
phantom-like behavior in a DGP-inspired scenario that curvature
effects are taken into account by incorporation of the Gauss-Bonnet
term in the brane part of the action. Although the Gauss-Bonnet
invariant essentially should be considered in the bulk action, we
can introduce it on the brane by considering a coupling between this
invariant and a scalar field on the brane. In this viewpoint, one
can consider possible modification of the induced gravity on the
brane by incorporation of higher order curvature terms such as the
Gauss-Bonnet type terms. Then, we have investigated the phantom-like
nature of this scenario with details. By the phantom-like behavior,
we mean an effective energy density which is positive and grows with
time and its equation of state parameter stays always less than
$-1$, that is, the effective equation of state parameter is in the
phantom region of the parameter space ( $\omega_{eff}^{DE}<-1$ ). In
this paper, firstly we have considered the normal branch of the
DGP-inspired Scalar-Gauss-Bonnet gravity ( which is ghost free) and
we have explored the cosmological implications of the model.
Especially, we have shown that this branch, which is not
self-accelerating in a pure DGP setup, can account for a
self-accelerating behavior at the late times due to the Gauss-Bonnet
curvature effect. As another important outcome of this setup, we
have shown that an effective phantom-like behavior can be realized
without need to introduce phantom fields neither on the brane nor in
the bulk. The effective equation of state parameter of the model has
a smooth crossing of the cosmological constant line at the near past
and in the same way as observations indicate. In the last stage, we
have considered a DGP-inspired model with modified-GB term in the
brane action. In this case, the brane part of the action is
dynamically equivalent to the Scalar-Gauss-Bonnet gravity with
vanishing kinetic-energy term. We have shown that this scenario can
also realize a phantom-like behavior by adopting cosmologically
viable ansatz. The phantom-like prescription is realized in this
setup for appropriate subspaces of the model parameter space ( for
instance with $-0.25\leq\beta\leq0.05$). It is important to note
that in the absence of the induced gravity effect, an effective
phantom phase has been obtained with negative powers of the scale
factor, that is, a phantom ansatz with $h_{0}<0$ in
$a=a_{0}(t_{s}-t)^{h_{0}}$. Here we have shown that the presence of
an induced gravity term on the brane leads to an effective
phantom-like behavior with a positive values of $h_{0}$ which we
call it a quintessence ansatz. Finally, we have shown that the
luminosity distance of the DGP-inspired scenario with curvature
correction is closer to the $\Lambda$CDM model than the $\Lambda$DGP
scenario. However, this result is dependent on the choice of the
model parameter space. Since the $\Lambda$CDM has very good
agreement with observational data, our analysis shows that with a
suitable choice of the parameter space of the model, the
DGP-inspired $F(G,\phi)$ scenario has better agreement with recent
observation than $\Lambda$DGP. This result seems to be reasonable
since the DGP-inspired $F(G,\phi)$ scenario has wider parameter
space than the $\Lambda$DGP model.

\end{document}